\documentclass[doublecol]{epl2}
\usepackage{amsmath}
\title{Resolving controversy of unusually high refractive index of tubulin}
\shorttitle{Title} 

\author{O. Krivosudsk\'{y}\inst{1} \and P. Dr\'{a}ber\inst{2} \and M. Cifra\inst{1}}
\shortauthor{F. Author \etal}

\institute{                    
  \inst{1} Institute of Photonics and Electronics, The Czech Academy of Sciences, Prague, Czechia\\
  \inst{2} Institute of Molecular Genetics, The Czech Academy of Sciences, Prague, Czechia
}

\abstract{Refractive index of tubulin is an important parameter underlying fundamental electromagnetic and biophysical properties of microtubules - protein fibers essential for several cell functions including cell division. Yet, the only experimental data available in the current literature show values of tubulin refractive index (n\,=\,2.36\,-\,2.90) which are much higher than established theories predict based on the weighted contribution of the polarizability of individual amino acids constituting the protein. To resolve this controversy, we report here modeling and rigorous experimental analysis of refractive index of purified tubulin dimer. Our experimental data revealed that the refractive index of tubulin is n\,=\,1.64 at the wavelength 589\,nm and 25\,$^{\circ}{\rm C}$, that is much closer to the values predicted by established theories than the earlier experimental data provide.}

\begin{document}
\maketitle
\section{Introduction}
Refractive index ($n$) is a fundamental material property which determines the interaction of light with a material. Microtubules, assembled from $\alpha\beta$-tubulin heterodimers, represent active biopolymer material which participates in essential cellular functions. $\alpha$- and $\beta$-tubulins are highly conserved and consist of isotypes encoded by different genes\,\cite{Luduena1}. Knowledge of tubulin refractive index is essential not only for accurate interpretation of results from novel cytoskeleton label-free imaging methods\,\cite{Bon}, construction and optical characterization of hybrid biological microtubule based nanodevices\,\cite{korten2010}, but since it is directly related to dielectric constant $\epsilon$ ($\epsilon$\,=\,n$^2$) it is also crucial for accurate electrostatic and coarse grained molecular modeling of microtubules\,\cite{baker2001, deriu2010} and design of di/electric manipulation techniques\,\cite{vanDenHeuvel2006}.
Well established theories predict refractive index of most proteins typically in the range $n$\,=\,1.5\,-\,1.7\,\cite{Zhao}. However, the experimental values of tubulin and microtubule refractive index available until now in the literature are in the range of $n$\,=\,2.36\,-\,2.90\,\cite{mershin3,Mershin,Bon}. Multiple studies\,\cite{schoutens2005, craddock2009, craddock2014} which employed this high value of tubulin refractive index have already been performed. To resolve this discrepancy and to provide solid data for assessing several related\,-\,proposed or observed\,-\,controversial quantum and electromagnetic properties of microtubules\,\cite{sahu2013, hameroff2014}, we carried out theoretical analysis of the refractive index of tubulin based on its primary structure and performed well characterized experiments to obtain refractive index increment ($dn/dc$) and refractive index of both unpolymerized and polymerized tubulin.

\section{Theory}
\begin{figure}[ht!]
\centering{\includegraphics{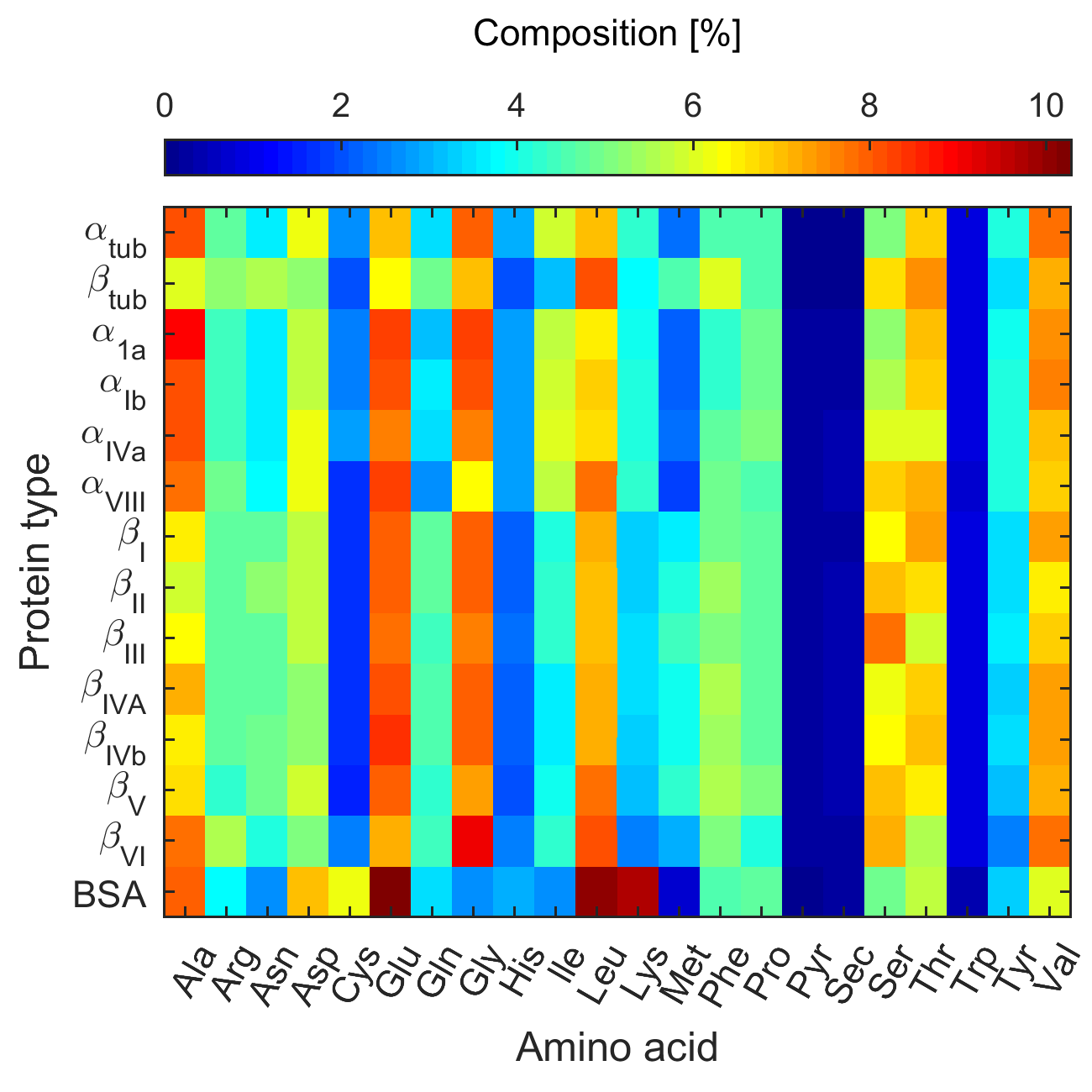}}
\caption{Amino acid composition of individual tubulin isotypes used in theoretical calculation of tubulin refractive index given by RCSB Protein Data Bank and UniProt database. Color encodes the percentual contribution of individual amino acids in tubulin molecule and its isotypes.}\label{fig:Histo}
\end{figure}

\begin{figure}[h!]
\centering{\includegraphics{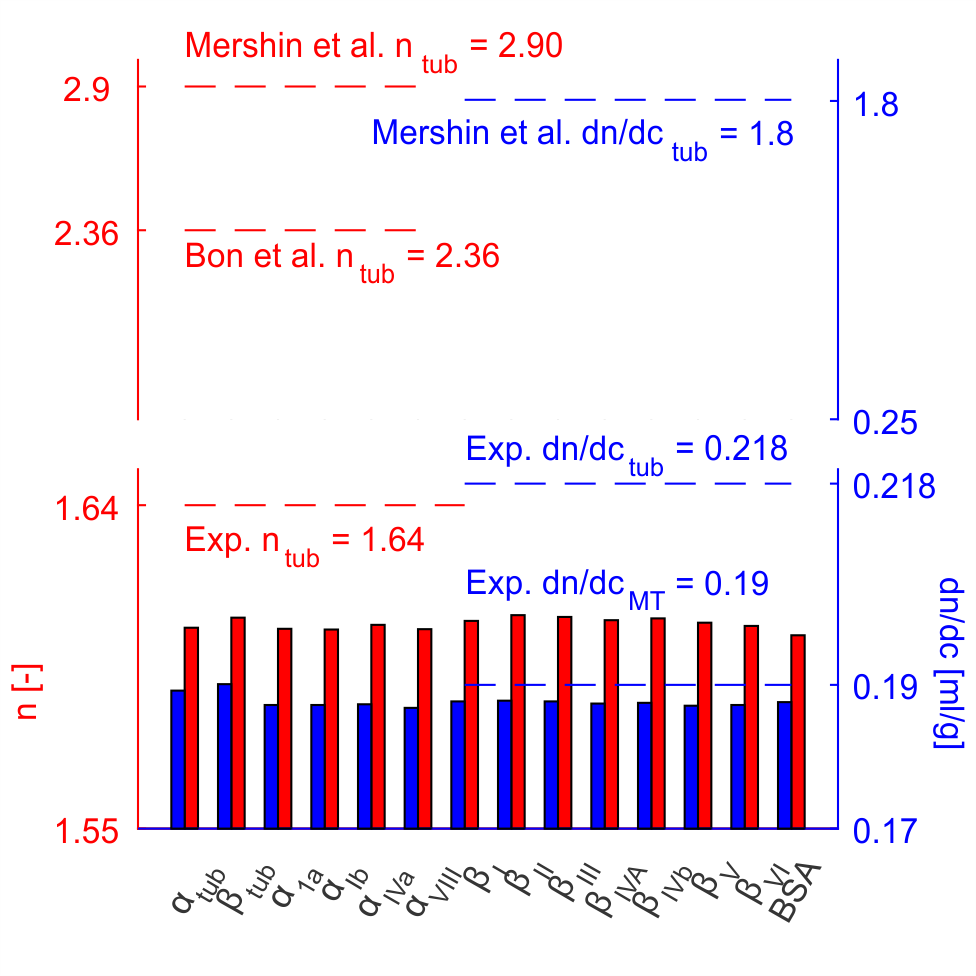}}
\caption{Bars represent theoretical values of refractive index (red bars) and $dn/dc$ (blue bars) of tubulin and bovine serum albumin. Dashed line are our experimental data (labeled as \textit{Exp.}) vs. earlier experimental data. Our data were experimentally obtained and theoretically calculated for the wavelength 589.3\,nm and\,25\,$^{\circ}{\rm C}$. The values from Mershin {\it et al.} \cite{Mershin,mershin3} are at 760\,nm and\,26\,$^{\circ}{\rm C}$ and from Bon {\it et al.} \cite{Bon} for 527\,nm and room temperature.}\label{fig:Histo2}
\end{figure}

Our theoretical analysis of refractive index of tubulin is based on the contribution of the polarizability of individual amino acids which constitute the tubulin isotypes. Number of amino acids contained in $\alpha\beta$\,-\,tubulin heterodimer was obtained from two different models. The first model is derived from a protein sequences of porcine tubulin (\textit{Sus scrofa}). In mammals, $\beta$\,-\,tubulin comprises 7 isotypes and $\alpha$\,-\,tubulin comprises 9 isotypes, each of which is a product encoded by distinct gene\,\cite{Luduena1,Luduena2,Luduena3}. We included several $\alpha$\,-\,tubulin isotypes which are found in porcine tubulin. Overview of used tubulin isotypes with UniProt database identifiers (ID) is shown in Tab.\,\ref{tab:tab1}. Detailed composition is displayed in Fig.\,\ref{fig:Histo}). Composition of second tubulin model is based on experimentally determined crystallographic structure  1TUB\,\cite{Nogales1998} obtained from RCSB protein database.\\
Refractive index of protein ($n_p$) can be obtained as a variation of Lorentz\,-\,Lorenz formula which is given as a ratio
\begin{equation}\label{eq:refra}
n_p=\sqrt{\frac{2R_p+\bar{v_p}}{\bar{v_p}-R_p}}
\end{equation}
where $R_p$ is the refraction per gram of protein given by an averaged contributions of its individual amino acids $R_{a}$\,\cite{McMeekin,McMeekin2} and their molecular masses $M_{a}$ in the following way
\begin{equation}\label{eq:R_p}
R_p=\frac{\sum_{a}{R_{a}M_{a}}}{\sum_{a}{M_{a}}}
\end{equation}
and protein partial specific volume $\bar{v}_p$ which can be also estimated as a weighted average of specific volumes ($\bar{v}_a$) of individual amino acids\,\cite{Durchschlag,Zamyatnin}
\begin{equation}\label{eq:v_p}
\bar{v}_p=\frac{\sum_{a}\bar{v}_{a}M_{a}}{\sum_{a}M_{a}}
\end{equation}
Based on the theoretical prediction of protein refractive index (Eq.\,\ref{eq:refra}) and a mixture rule represented by Wiener equation\,\cite{Heller} the refractive index increment can be estimated by using the following equation
\begin{equation}\label{eq:Mix}
\frac{dn}{dc}=\frac{3}{2}\bar{v}_p n_{buffer}\frac{n_p^2-n_{buffer}^2}{n_p^2+2n_{buffer}^2}
\end{equation}
with a buffer refractive index $n_{buffer}$.\\
Based on the values of refraction per gram and specific volumes of individual amino acids available in literature\,\cite{McMeekin, McMeekin2, Durchschlag, Zamyatnin} and tubulin composition, we predicted $dn/dc$ and refractive index of the tubulin isotypes as well as of the bovine serum albumin (BSA) as a reference protein. See Fig.~\ref{fig:Histo2} and supp. material for more details.

\begin{table}[ht!]
\begin{center}
\begin{tabular}{ccc}
 \hline
  Isotype & Gene name & ID \\
  		  && UniPortKB \\
 \hline \hline
 	\text{$\alpha1A$}&$TUBA1A$&P02550\\
    \text{$\alpha1B$}&$TUBA1B$&Q2XVP4\\
    \text{$\alpha4A$}&$TUBA4B$&F2Z5S8\\
    \text{$\alpha8$}&$TUBA8$&I3LDR2\\
    \text{$\beta{I}$}&$TUBB$&Q767L7\\
    \text{$\beta{II}$}&$TUBB2B$&F2Z5B2\\
    \text{$\beta{III}$}&$TUBB3$&F1S6M7\\
    \text{$\beta{IVa}$}&$TUBB4A$&F2Z5K5\\
    \text{$\beta{IVb}$}&$TUBB4B$&F2Z571\\
    \text{$\beta{V}$}&$TUBB6$&I3LBV1\\
    \text{$\beta{VI}$}&$TUBB1$&A5GFX6\\    
\hline  
   &  & ID \\
  		  && RCSB database \\
\hline      \hline    
	\text{$\alpha$}\,-\,tubulin chain & $1TUB:A$ & GI:3745821 \\
	\text{$\beta$}\,-\,tubulin chain & $1TUB:B$ & GI:3745822\\
Bovine serum & $4F5S:A,B$ & P02769\\
albumin&&\\                                   
\hline
	\end{tabular}
\caption{Overview of \emph{Sus scrofa} tubulin isotypes and BSA.}{\label{tab:tab1}}
\end{center}
\end{table}

\begin{figure}[ht!]
\centering{\includegraphics{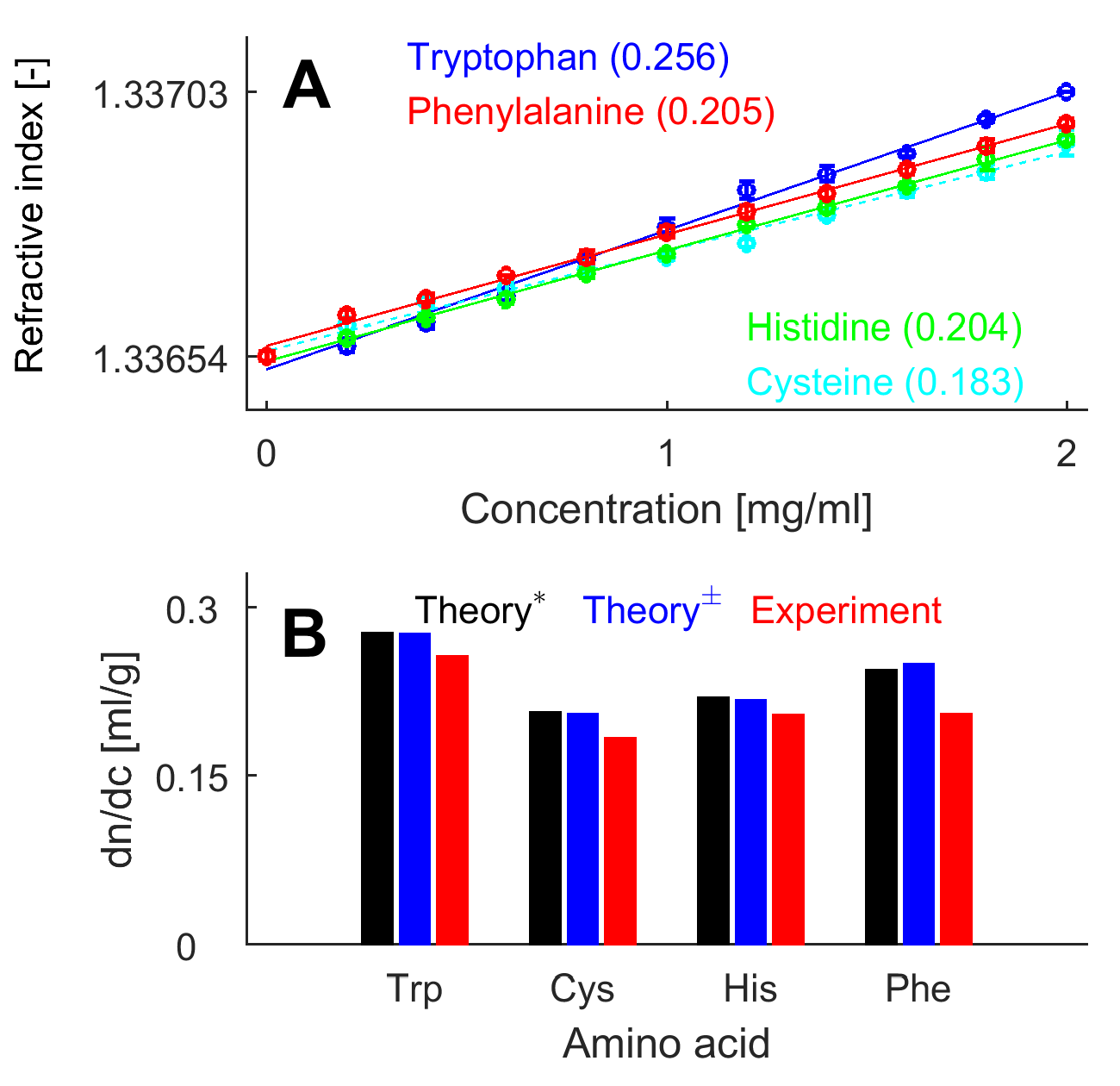}}
\caption{A) Refractive index $n$ of amino acid solutions. Data points were fitted by first order polynomial (linear regression was performed). Slope of the curve represents refractive index increment of given solution. B) Predicted theoretical and experimental data (red) of amino-acids. Theory$^\ast$ (black) - Data were predicted at 589\,nm for hypothetical polypeptide in water with 150\,mM NaCl\cite{Zhao}. Theory$^\pm$ (blue) - Data calculation is based on evaluation of Eq\,\eqref{eq:Mix} for amino acid buffer (BRB80) solution with specific volumes and refraction of individual amino acids\,\cite{McMeekin, McMeekin2, Durchschlag, Zamyatnin}.}\label{fig:Concentration}
\end{figure}

To measure refractive index, we used a critical-angle dispersion-refractometer with thermal regulation (DSR\,-\,$\lambda$ SCHMIDT\,+\,HAENSCH) in all experiments. At first, we experimentally analyzed $dn/dc$ of selected amino acids (see Fig.\,\ref{fig:Concentration}\,A) and thereby verified that our method yields values consistent with the literature\,\cite{McMeekin, McMeekin2}. All refractive index measurements were carried out at 589\,nm and sample temperature 25\,$^{\circ}{\rm C}$ since the values of amino acid $dn/dc$ from literature were also at this wavelength and temperature.

\begin{figure}[ht!]
\centering{\includegraphics{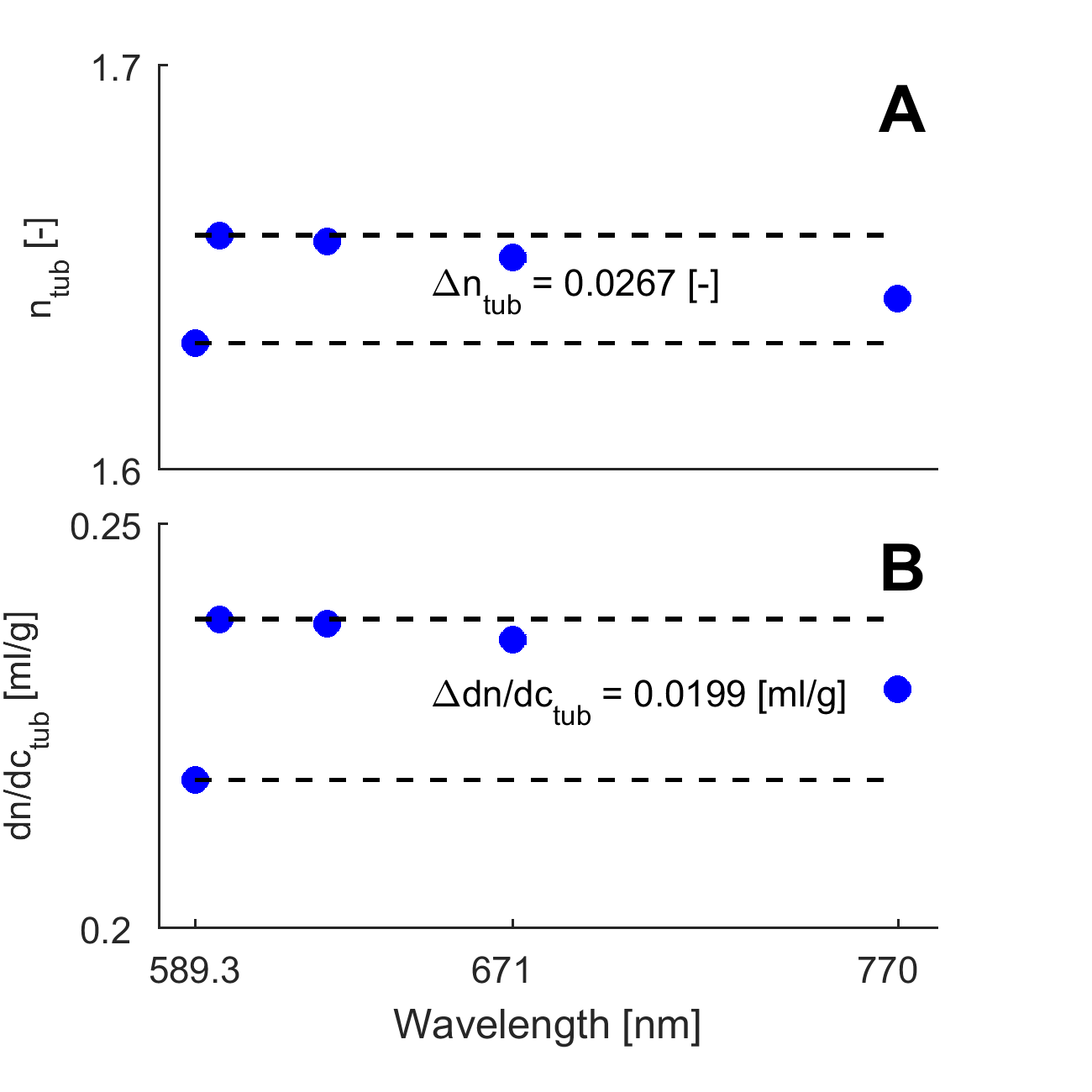}}
\caption{A) Refractive index of tubulin at different wavelengths calculated from B) refractive index increments.} \label{fig:Wavelength}
\end{figure}

Additionally, this temperature also prevents spontaneous polymerization of the tubulin in our conditions. All amino acids, Tubulin ($>$\,99\,\% purity, \#\,T238P, Cytoskeleton, Inc.) and BSA ($\geq$98\% purity, A3803, Sigma-Aldrich) were stored in the BRB80 buffer (80\,mM PIPES, 2\,mM MgCl2, 0.5\,mM EGTA, 1\,mM GTP, 6.9\,pH, KOH). For tubulin and BSA, refractive index of their concentration series from 0 to 2 mg/ml (11 steps, each with 50 $\mu$l sample volume) has been measured (each series took 75 min, with tubulin run in independent triplicate) to obtain $dn/dc$. At last, refractive index increment of polymerized tubulin was measured. Tubulin was polymerized in BRB80 buffer with adding taxol stepwise at 2, 20 and 200\,$\mu$M steps at 37\,$^{\circ}{\rm C}$\,\cite{Mitchison}.
Results in Fig.\,\ref{fig:Concentration}\,A indicate that amino acid $dn/dc$ are independent on concentration. A least-square fit linear regression yields straight lines, and $dn/dc$ of amino acids did not deviate significantly from established theoretical values, for details see supp. material.
We obtained refractive index of proteins from their measured $dn/dc$ (Fig.\,\ref{fig:Histo2}) following the method outlined in\,\cite{Mershin}. In brief, assuming the contribution from all the buffer components we can write refractive index of tubulin as
\begin{equation}\label{eq:index}
n_{tub}=\frac{n_{sol}-(1-\chi_{p})n_{buffer}}{\chi_{p}}
\end{equation}
where $\chi_{p}=C_{tub}/\rho$ is the protein mass fraction with density $\rho$ and refraction index of the solution $n_{sol}$ which is based on the refractive index increment ($dn/dc$) in following way
\begin{equation}
n_{sol}=\frac{dn}{dc}C_{tub}+n_{buffer}
\end{equation}
As a tubulin concentration $C_{tub}$\,=\,1.60 mg/ml was used with the protein density $\rho$\,=\,1.41\,g/ml. Given by a experimental data of tubulin increment (see Fig.\,\ref{fig:Histo2}) solution refractive index follows as
\begin{equation}
n_{sol}=2.18\times10^{-4}C_{tub}+1.33654
\end{equation}
and by applying Eq.\,\ref{eq:index}, the refractive index of tubulin was found to have the value
\begin{equation}
n_{tub}=1.64\pm0.02@\,589\,nm\,@\,25\,^{\circ}{\rm C}
\end{equation}
By using the same procedure with bovine serum albumin, $dn/dc_{BSA}$\,=\,0.179 ml/g was measured which results in the refractive index $n_{BSA}$=1.59\,$\pm$\,0.02, which is in exact agreement with the theoretical predictions. For microtubules, we experimentally obtained value $dn/dc_{MT}$\,=\,0.19 ml/g which would correspond to $n_{MT}$\,=\,1.60. However, the Lorentz-Lorentz formula used to calculate polarizability and refractive index based on the $dn/dc$ assumes spherical approximation of the particle\,\cite{born2000}[p.89-92] which obviously is not valid for the microtubules. The small discrepancies between the theory and experiment for $n_{tub}$ and  $dn/dc_{tub}$ can be due to (i) the deviation of tubulin dimer shape from a spherical one assumed in the theory, (ii) small concentration uncertainties, (iii) effects of the buffer on $dn/dc$\,\cite{ball1998} or (iv) additional polarization effects in tubulin due to its large dipole moment\,\cite{tuszynski2006,tuszynski12}.\\
The high-frequency polarizability ($\alpha_{tub}$) of tubulin can be also evaluated by using well known Clausiu-Mossotti relationship in the following form:
\begin{equation}
\alpha_{tub}=\frac{3\varepsilon_0}{N}\frac{n^2_{tub}-1}{n^2_{tub}+2}
\end{equation}
which with the calculated number density of molecules $N$\,=\,8.7$\times$10$^{21}$\,molecules per m$^3$ at given concentration (C$_{tub}$) is equal to $\alpha_{tub}$\,=\,1.1$\times$10$^{-33}$\,Cm$^2$V$^{-1}$.\\
What effect caused the high values of detected refractive index in the earlier experimental works\,\cite{Mershin,mershin3,Bon} is unclear. In case of \cite{Mershin,mershin3}, the $n_{tub}$\,=\,2.90\,$\pm$\,0.10 was obtained using both by surface plasmon resonance sensing as well as refractometry technique similar to ours. While authors of  \cite{Mershin,mershin3} used different wavelength (760\,nm) than us (589\,nm), we found from our experiments only a very small  difference of $n$ ($\Delta n = n_{tub@770}$\,-\,$n_{tub@589}$\,$<$\,0.011 , see Fig.\,\ref{fig:Wavelength} for wavelength dependence) which does not explain the discrepancy. 
In the newer work\,\cite{Bon}, refractive index of single microtubules was extracted using a quantitative phase imaging based microscopy technique from fixed fluorescently labeled cells with a rather large error ($n_{MT}\,=\,2.36\,\pm\,0.6$) - the cause of the high value of $n$ is not entirely clear but might be due to the assumptions in the model required to extract the $n$ from their experimental data.
\section{Conclusion}
To summarize, we provide experimental data supported by theoretical modeling which reveal that the refractive index of unpolymerized and polymerized tubulin is in the range $n_{tub}$\,=\,1.6\,-\,1.64, yielding high frequency dielectric constant $\epsilon$\,=\,2.56\,-\,2.69. This value provides lower dielectric screening within the tubulin than previously thought. Lower screening enables larger interaction distance of electrostatic forces within microtubules than previously considered based on the higher values of $\epsilon$\,\cite{schoutens2005}. Further implications are for the exciton energy transfer within the tubulin; lower $n$ enhances electronic dipole-dipole coupling due its 1/$n^{4}$ dependence\,\cite{andrews2011} suggesting resonant energy transfer on a longer range than previously estimated\,\cite{craddock2014}. We believe our data will enable more accurate tubulin and microtubule characterization and modeling much beyond the few examples we illustrated.

\section{Acknowledgments}
The research presented in this paper was supported by the Czech Science Foundation, projects no. P102/15-17102S and P302/16-25159S and by Institutional Research Support (RVO 68378050). Authors participate in COST Action BM1309 and bilateral exchange project between Czech and Slovak Academies of Sciences, no. SAV-15-22. 

\section{Authors contributions}
OK conceived the experiments and modeling and drafted the paper, PD provided the experimental material and important
methodical suggestions, MC designed the research and co-wrote the paper.


\end{document}